\documentclass[aps,prl,twocolumn,showpacs,superscriptaddress]{revtex4-1}  
\usepackage{graphicx}  
\usepackage{dcolumn}   
\usepackage{bm}        
\usepackage{amssymb}   
\usepackage{amsmath}   
\usepackage{color}

\begin{document}

\title{Identifying structural flow defects in disordered solids using machine learning methods}
\author{E. D. Cubuk}
\email{cubuk@fas.harvard.edu}
\affiliation{Department of Physics and School of Engineering and Applied Sciences, Harvard University, Cambridge, Massachusetts 02138, USA}
\author{S. S. Schoenholz (Equal contribution)}
\email{schsam@sas.upenn.edu}
\affiliation{Department of Physics, University of Pennsylvania, Philadelphia, Pennsylvania 19104, USA}
\author{J. M. Rieser}
\affiliation{Department of Physics, University of Pennsylvania, Philadelphia, Pennsylvania 19104, USA}
\author{B. D. Malone}
\affiliation{Department of Physics and School of Engineering and Applied Sciences, Harvard University, Cambridge, Massachusetts 02138, USA}
\author{J. Rottler}
\affiliation{Department of Physics and Astronomy, University of British Columbia, Vancouver, BC V6T1Z4, Canada}
\author{D. J. Durian}
\affiliation{Department of Physics, University of Pennsylvania, Philadelphia, Pennsylvania 19104, USA}
\author{E. Kaxiras}
\affiliation{Department of Physics and School of Engineering and Applied Sciences, Harvard University, Cambridge, Massachusetts 02138, USA}
\author{A. J. Liu}
\affiliation{Department of Physics, University of Pennsylvania, Philadelphia, Pennsylvania 19104, USA}

\begin{abstract}
We use machine learning methods on local structure to identify flow defects -- or regions susceptible to rearrangement -- in jammed and glassy systems. We apply this method successfully to two disparate systems: a two dimensional experimental realization of a granular pillar under compression, and a Lennard-Jones glass in both two and three dimensions above and below its glass transition temperature. We also identify characteristics of flow defects that differentiate them from the rest of the sample. Our results show it is possible to discern subtle structural features responsible for heterogeneous dynamics observed across a broad range of disordered materials.
\end{abstract}

\pacs{83.50.-v, 62.20.F-, 61.43.-j}

\maketitle

All solids flow at high enough applied stress and melt at high enough temperature. Crystalline solids flow~\cite{taylor34} and  premelt~\cite{alsayed05} via localized particle rearrangements that occur preferentially at structural defects known as dislocations.  The population of dislocations therefore controls both how crystalline solids flow and how they melt.  In disordered solids, it has long been hypothesized that localized particle rearrangements~\cite{argonkuo79} induced by stress or temperature also occur at localized flow defects~\cite{argon79,falklanger98,bocquet2009}.  Like dislocations in crystals~\cite{rottler14}, flow defects in disordered solids are particularly effective in scattering sound waves, so analyses of the low-frequency vibrational modes~\cite{brito07} have been used successfully to demonstrate the existence of localized flow defects ~\cite{widmer08,cooper09,candelier10,manning11,chen11,hocky13,chen13,rottler14,schoenholz14,jack14}.  However, all attempts to identify flow defects~\cite{gilman75,chaudhari79} directly from the structure, without using knowledge of the inter-particle interactions, have failed~\cite{gilman75,chaudhari79}. 
Likewise, in supercooled liquids, purely structural measures correlate only weakly with kinetic heterogeneities~\cite{jack14}, although correlations between structure and dynamics have been established indirectly~\cite{cooper04,cooper07,coslovich06,coslovich07,malins13}.  

Here we introduce a novel application of machine learning (ML) methods to identify ``soft" particles that are susceptible to rearrangement or, equivalently, that belong to flow defects, from the local structural geometry alone. We apply the method to two very different systems--an experimental frictional granular packing under uniaxial compression and a model thermal Lennard-Jones glass in both two and three dimensions. The analysis of granular packing shows that our method succeeds even when previous methods based on vibrational modes~\cite{manning11} are inapplicable.  The results for Lennard-Jones systems show that the correlation between structure and irreversible rearrangements does not degrade with increasing temperature, even above the dynamical glass transition, and is equally strong in two and three dimensions.
Finally, we exploit the method to discover which structural properties distinguish soft particles from the rest of the system, and to understand why previous attempts to identify them by structural analysis have failed.

Physically-motivated quantities such as free volume or bond orientational order correlate with flow defects~\cite{manning11}, but are insufficient to identify them \textit{a priori}. We introduce a large set of quantities that are each less descriptive, but when used as a group provide a more complete and unbiased description of local structure.  These quantities have been used to represent the potential energy landscape of complex materials from quantum mechanical calculations~\cite{behlerPRL_orig}. For a system composed of multiple species of particles, we define two families of structure functions for each particle $i$,
\begin{align}
\label{eq:symfunc1}
&G_{Y}^{X}(i;\mu) = \sum_j e^{-(R_{ij}-\mu)^2/L^2}\\
\label{eq:symfunc2}
&\Psi_{YZ}^{X}(i;\xi,\lambda,\zeta) = \sum_j\sum_k e^{-(R_{ij}^2 + R_{ik}^2 + R_{jk}^2)/\xi^2}(1+\lambda\cos\theta_{ijk})^\zeta
\end{align}
where $R_{ij}$ is the distance between particles $i$ and $j$; $\theta_{ijk}$ is the angle between particles $i$, $j$ and $k$;  $L$, $\mu$, $\xi$, $\lambda$, and $\zeta$ are constants; $X$,$Y$,$Z$ are labels that identify the different species of particles in the system, with the correspondence $i\leftrightarrow X$, $j\leftrightarrow Y$, $k\leftrightarrow Z$. By using many different values of the constants $\mu$, $\xi$, $\lambda$, and $\zeta$ we generate many structure functions in each family that characterize different aspects of a particle's local configuration; for a list, see supplementary information. 
The first family of structure functions $G$ characterizes radial density properties of the neighborhood, while the second family, $\Psi$, characterizes bond orientation properties. The sums are taken over particle pairs whose distance is within a large cutoff $R_c^S$. Our results are qualitatively insensitive to the choice of $R_c^S$ as long as it includes several neighbor shells.

Having characterized local structure through $G_{Y}^{X}(i;\mu)$ and $\Psi_{YZ}^{X}(i;\xi,\lambda,\zeta)$, we introduce a method to infer from this information the location of flow defects in disordered solids. Generically, we begin with a set of $N$ particles to be classified as ``soft'' or ``hard.'' Each particle is described by a set of $M$ variables derived from the two families of functions $G_{Y}^{X}$ and $\Psi_{YZ}^{X}$ by varying the constants $\mu$, $\xi$, $\lambda$, and $\zeta$ (here $M = 160$); this is represented by the set of vectors $\{\bm F_1,\cdots,\bm F_N\}$, where $\bm F_i$ constitutes an embedding of the local environment of a particle $i$, constructed at a time $t_i$, in $\mathbb R^M$. We select at random a subset of $n$ of these particles (the ``training set'') and categorize them \textit{a posteriori} as being soft if they rearrange (the details of which will be discussed below) between time $t_i$, when the structure is characterized, and time $t_i+\Delta t$. Otherwise the particles are labeled as hard. 

The next step is to use the particles in the training set, already classified as soft or hard, to construct a scheme to identify other particles as soft or hard.   We use the support vector machine (SVM) method~\cite{SVM}, which constructs a hyperplane in $\mathbb R^M$ that best separates soft particles from hard ones. Once this hyperplane has been established for the training set, the rest of the particles (and any particles from similar systems) may be classified according to whether their local structures place them on soft or hard sides of the hyperplane. Generically, no exact separation is possible so the SVM method is adapted to penalize particles whose classification is incorrect; the degree of penalty is controlled by a parameter $C$ where larger values of $C$ allow for fewer incorrect classifications. We find that the quality of our classifications is insensitive to $C$ for $C>0.1$. The SVM algorithm was implemented using the LIBSVM package~\cite{libsvm}.

To identify rearrangements we calculate, for each particle $i$, the widely-used quantity~\cite{falklanger98} 
\begin{equation}
D^2_{\text{min}}(i)\equiv \min_\Lambda \left\{ \frac{1}{z}\sum_j (R_{ij}(t+\Delta t)-\Lambda R_{ij}(t))^2 \right\} ,
\end{equation} 
which characterizes the magnitude of non-affine displacement during a time interval $\Delta t$. Here the sum runs over neighbors $j$ within a distance of $R^D_c$ of particle $i$, $R_{ij}$ is the center-center displacement between particles $i$ and $j$, $z$ is the number of neighbors within $R^D_c$.  The quantity is minimized over choices of the local strain tensor $\Lambda$. We find $D^2_{\text{min}}$ is insensitive to the choice of $R^D_c$ and $\Delta t$ so long as $R^D_c$ is large enough to capture the particles local neighborhood and $\Delta t$ is longer than the ballistic timescale. A particle is said to have undergone a rearrangement if $D^2_{\text{min}} \ge D^{2}_{\text{min,0}}$. We choose $D^2_{\text{min,0}}$ such that approximately $0.15\%$ of the particles from each species in each system has gone through a rearrangement although the results depend little on the specific choice of $D^2_{\text{min},0}$.

\begin{figure}[!h]
\centering
\includegraphics[width=0.35\linewidth]{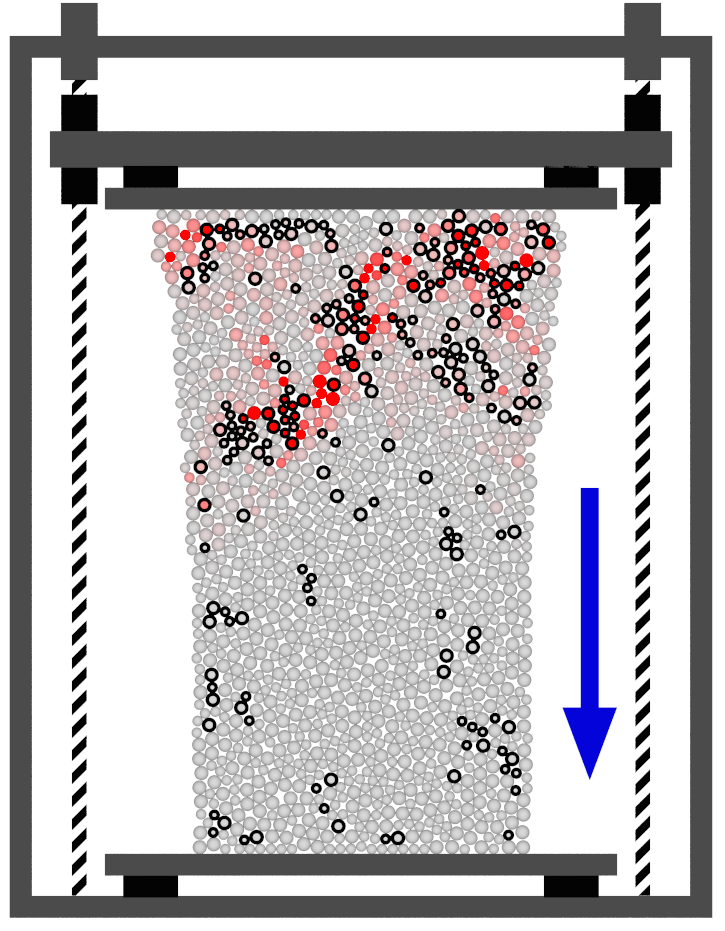}
\hspace{1pc}
\includegraphics[width=0.45\linewidth]{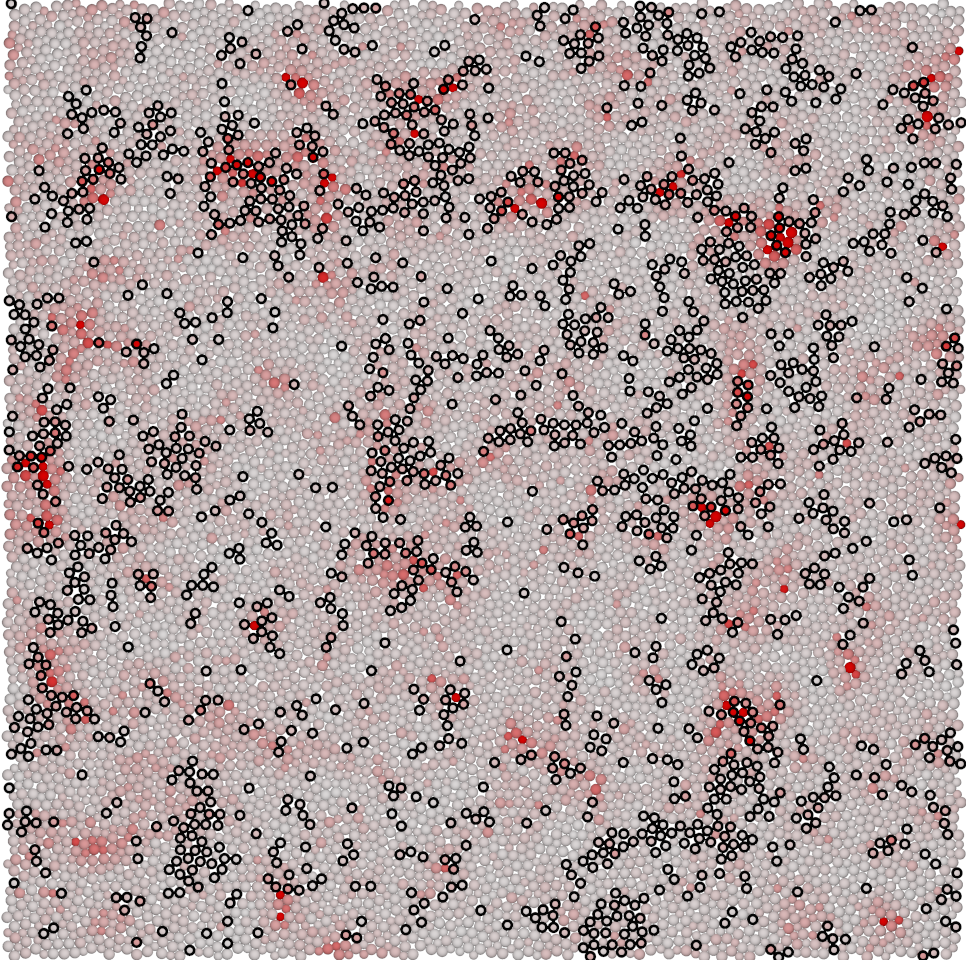}
\caption{Snapshot configurations of the two systems studied. Particles are colored gray to red according to their $D^2_{\text{min}}$ value. Particles identified as soft by the SVM are outlined in black. (a) A snapshot of the pillar system. Compression occurs in the direction indicated. (b) A snapshot of the $d=2$ sheared, thermal Lennard-Jones system.}
\label{fig:system}
\centering
\end{figure}

We first test our approach on an experimental system of two-dimensional ($d=2$) ``pillars" of particles. A bidisperse pillar made up of grains (plastic cylinders resting upright on a horizontal substrate) is situated between two plates in a custom-built apparatus.  The bottom plate is fixed and the top plate is driven into the pillar at a constant speed of $v_0=0.085$ mm/s. The pillars are composed of a bidisperse mixture of approximately 1500 rigid grains with size ratio 3:4 and the large particles having a radius of $d_{AA}=0.3175$cm. These particles have elastic and frictional interactions with each other, as well as frictional interactions with the substrate, making the identification of flow defects using vibrational modes impossible. A camera is mounted above and captures images at 7 Hz throughout the compression. 

We construct our training set from compression experiments performed on ten different pillars. We select 6,000 particles at random from the entire duration of the experiments that undergo a rearrangement in the next 1.43 seconds and an equal number of particles that do not. To identify rearrangements we calculate $D^2_{\text{min}}$ with $R^D_c = 1.5\cdot d_{AA}$ and $D^2_{\text{min,0}} = 0.25d_{AA}^2$. Compression of the mechanical pillar from the top only affects particles above a certain ``front'' that starts at the top and advances towards the bottom with time. Our training set contains only particles within this activated front.  Particles in a horizontal slice between $y$ and $y+\delta y$ are said to be within the activated front if the average speed of particles in the slice exceeds $v_{\text{thresh}} \sim 0.04$ mm/s.  

As a second test, we apply our approach to a model glass in both $d=2$ and $d=3$. We study a 65:35 binary Lennard-Jones (LJ) mixture with $\sigma_{AA} = 1.0$, $\sigma_{AB} = 0.88$, $\sigma_{BB} = 0.8$, $\epsilon_{AA} = 1.0$, $\epsilon_{AB} = 1.5$, and $\epsilon_{BB} = 0.5$~\cite{bruning09}. The LJ potential is cut off at $2.5\sigma_{AA}$ and smoothed so that both first and second derivatives go continuously to zero at the cutoff. The natural units for the simulation are $\sigma_{AA}$ for distances, $\epsilon_{AA}$ for energies, and $\tau = \sqrt{m\sigma_{AA}^2/\epsilon_{AA}}$ for times. 
We perform molecular dynamics simulations using LAMMPS~\cite{LAMMPS} with a time step of $5\times 10^{-3}\tau$ at density $\rho=1.2$, using a Nos\'e-Hoover thermostat with a time constant of $\tau$. Here, temperature is in units of $\epsilon_{AA}/k_B$, where $k_B$ is the Boltzmann constant.  In $d=2$, we consider a system of 10,000 particles at temperatures $T = 0.1, 0.2, 0.3,$ and $0.4$ at a single strain rate, $\dot \gamma=10^{-4}/\tau$. In all cases data was collected after allowing the system to reach steady state by shearing up to 20\% strain. In $d=3$, we use a collection of 30,000 particles at temperatures $T=0.4, 0.5, 0.6$ with no strain. The quiescent system has a glass transition at $T_G = 0.33$ in $d=2$ and $T_G=0.58$ in $d=3$~\cite{bruning09}. Therefore, in both dimensions we study the system both above and below its glass transition temperature.

At each temperature we construct training sets of 6,000 and 10,000 particles, in $d=2$ and 3, respectively, selected at random from the entire run that undergo a rearrangement in the following $\Delta t=2\tau$ units of time, and an equal number of particles that do not undergo a rearrangement. To identify rearrangements, we calculate $D^2_{\text{min}}$ with $R^D_c=2.5\sigma_{AA}$ to be the same as the range of the truncated LJ potential and $\Delta t = 2\tau$. In $d=3$ the $D^2_{\text{min}}$ distributions of the species A and species B particles differ significantly and so are treated separately throughout the analysis.

\begin{figure}[t]
\centering
\includegraphics[width=0.85\linewidth]{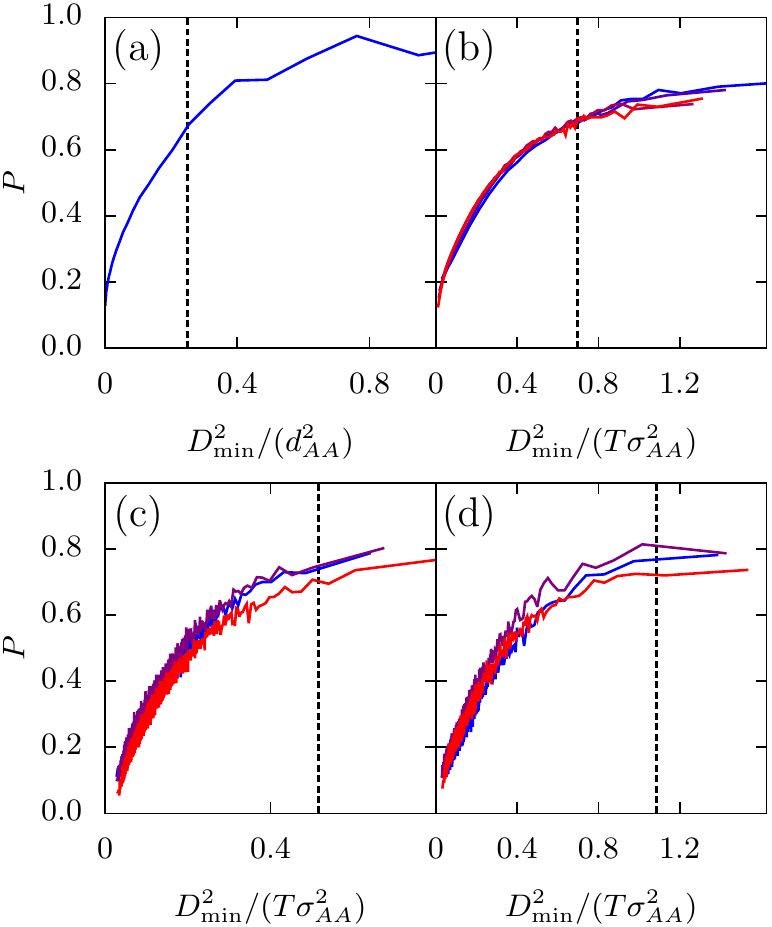}
\caption{Probability that a particle of a given $D^2_{\text{min}}$ value is soft. The vertical dashed lines are corresponding $D^2_{\text{min,0}}$ values. (a) The result for the pillar system, where $d_{AA}$ refers to the large grain diameter (since this is a granular system with macroscopic grains, thermal fluctuations are negligible). (b) The result of using an SVM trained at a temperature $T$ ($T = 0.1, 0.2, 0.3$ and $0.4$ shown in different colors ) to classify data at the same temperature for the $d=2$ LJ glass. (c,d) Results for species A and B, respectively, for the $d=3$ system at $T = 0.4, 0.5$ and $0.6$.}
\label{fig:probability}
\centering
\end{figure}

We now test our classifying hyperplane on the three systems outlined above. For each system we construct a test set of particles that were not used in training the SVM; for the pillar this test set consists of 100,000 particle environments from ten additional compression experiments. For the $d=2$ and $d=3$ LJ glasses we use $2\times 10^7$ and $75\times10^7$ unseen particle environments respectively. Fig.~\ref{fig:probability} (a)-(d) shows the probability, $P(D^2_{\text{min}})$, that a particle with an observed value of $D^2_{\text{min}}$ was identified as soft by our classification \textit{a priori} for each system. Fig~\ref{fig:probability} (c)-(d)  treat the particles of species A and species B separately for the $d=3$ LJ glass. In all cases we see that $P(D^2_{\text{min}})$ rises with increasing plastic activity, $D^2_{\text{min}}$. This implies that the particles identified as soft by the SVM are more likely to be involved in plastic flow.  For the granular pillar, 21\% of the particles are classified as soft, and these particles capture 80\% of the rearrangements.  For the $d=2$ and $d=3$ LJ systems, these numbers are 26\% and 73\%, and 24\% and 72\%, respectively, at the highest temperatures studied. Thus, we consistently find rearrangements to occur at soft particles about 3-4 times more frequently than if the soft particles were randomly chosen. Finally, Fig.~\ref{fig:probability}(b,c,d) show that $P(D^2_{\text{min}})$ collapses for different $T$ both for 2D and 3D systems, when $D^2_{\text{min}}$ is scaled by $T \sigma_{AA}^2$; this scaling arises since for particles not undergoing rearrangement, $D^2_{\text{min}}\sim\langle v^2\rangle\sim T$ by the equipartition theorem \cite{schoenholz14}.

Remarkably, our ability to identify soft particles does not decrease with increasing temperature or dimension. For the $d=2$ LJ system over the same temperature range, the accuracy of the vibrational mode method decreases by over 50\%~\cite{schoenholz14}.  The key difference between the two methods is that we construct local environments from the actual particle positions in snapshots of the thermal system, while soft particles from vibrational modes are extracted from particle positions in the inherent structures obtained by quenching to $T=0$.

\begin{figure}[!h]
\centering
\includegraphics[width=0.85\linewidth]{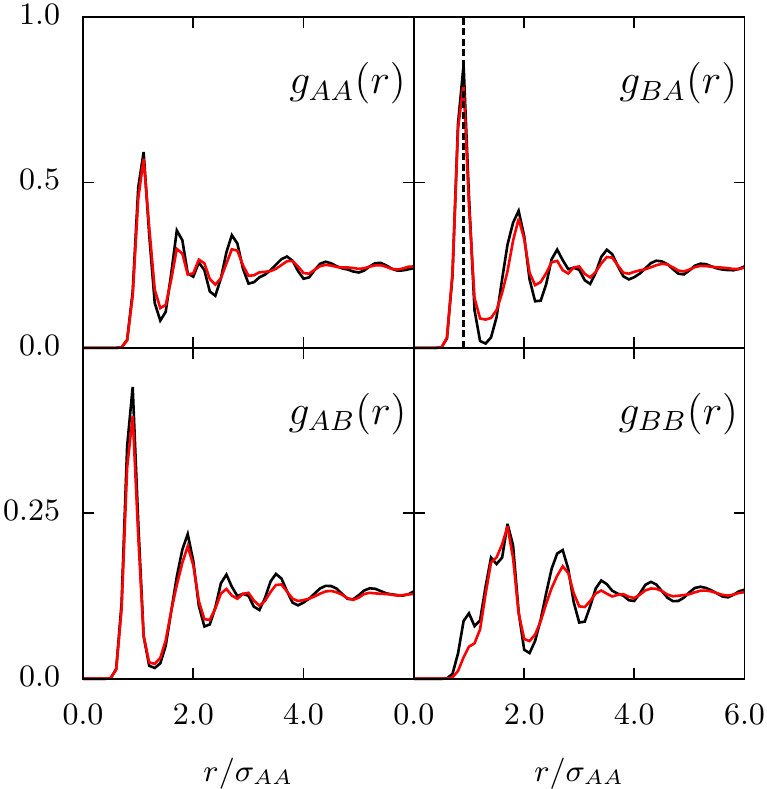}
\caption{Radial distribution functions averaged over hard (black lines) or soft (red lines) particles. $g_{AB}$ and $g_{BA}$ of soft particles are not equal to each other since they refer to different kinds of regions: neighbors of soft particles from species A and neighbors of soft particles from species B, respectively.}
\label{fig:physical1}
\centering
\end{figure}

\begin{figure}[!h]
\centering
\includegraphics[width=0.85\linewidth]{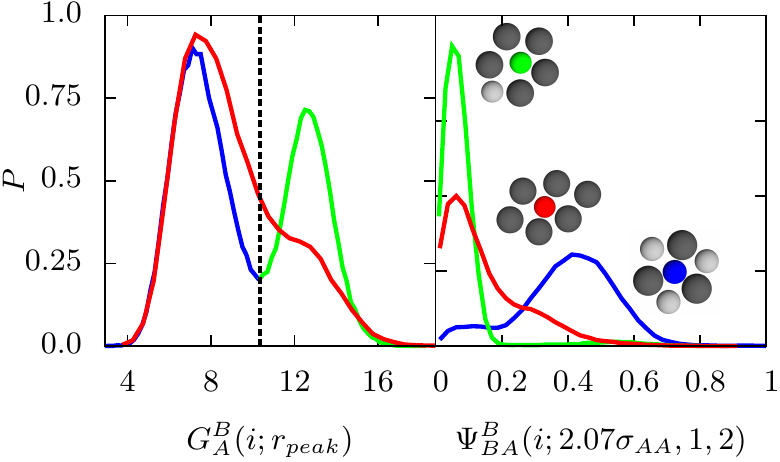}
\caption{(a) Distribution of $G_{B}^A(i;r_{\text{peak}}^{AB})$, proportional to the gaussian weighted density at $r_{\text{peak}}^{AB}$, for soft (red) and hard (blue/green) particles. $r_{\text{peak}}^{AB}$ corresponds to the first peak of $g_{AB}$ or $g_{BA}$. (b) Distribution of $\Psi^B_{AB}(i;2.07\sigma_{AA},1,2)$, proportional to the density of neighbors with small bond angles near a particle $i$, for soft (red) and hard (blue/green) particles. The inset shows examples of configurations with corresponding radial and bond orientation properties, where dark (light) gray neighbors are of species A (B).}
\label{fig:physical2}
\centering
\end{figure}

The hyperplane that divides soft from hard particles tells us which features of the local structural environment are important in distinguishing between the hard and soft particles. To illustrate this, we focus on the sheared $d=2$ LJ glass at $T = 0.1$.  The first set of structure functions, $G^X_{Y}(i;\mu)$ defined in Eq.~(\ref{eq:symfunc1}), has a familiar physical interpretation in terms of the radial distribution function: $g_{XY}(r) = \lim_{L \to 0} \langle G_{Y}^X(i;r)\rangle/2\pi r$. This family of structure functions is essentially a parameterization of a particle's local contribution to $g(r)$. In Fig.~\ref{fig:physical1} we show the approximations to various $g(r)$ constructed in this way, using $L=0.1 \sigma_{AA}$, for particles identified as hard (black lines) and soft (red lines). 

In all cases, we see that soft particles feature slightly lower peaks and higher troughs in $g(r)$. To see whether this difference is sufficient to identify soft particles, we expand our analysis beyond average values of $G^X_{Y}(i;\mu)$, to the \emph{distributions} of the different structure functions. In Fig.~\ref{fig:physical2}(a) we show the distribution of values of $G^B_{A}(i;r_{\text{peak}}^{AB})$ where $r_{\text{peak}}^{AB}$ is the location of the first peak of $g_{BA}(r)$, for soft particles (red) and hard particles (blue/green). While the distribution for soft particles features a single peak, that for hard particles is bimodal. This indicates the existence of (at least) two distinct populations of hard particles which we divide into two groups: one with $G^B_{A}(i;r_{\text{peak}}^{AB})/r < 1/2$ (blue) that we will call $H_0$-type, and one with $G^B_{A}(i;r_{\text{peak}}^{AB})/r > 1/2$ (green) that we will call $H_1$-type. Radial information therefore distinguishes between soft particles and $H_1$ hard particles but not between soft and $H_0$ hard particles.

We now consider the distribution of $\Psi^B_{AB}(i;2.07\sigma_{AA},1,2)$ for soft particles (red), $H_0$ hard particles (blue), and $H_1$ hard particles (green), shown in Fig. 3(c). Physically, $\Psi^B_{AB}(i;2.07r_{\text{peak}}^{BA},1,2)$ is large when there are many pairs of neighbors of the central particle that lie within a distance $\xi$ with small angles between them, such that one is of species A and one is of species B. The soft particles fall into a single category (one peak, red) while $H_0$-type and $H_1$-type hard particles--defined from radial information above--have very different distributions (blue and green peaks). Unlike before, here the soft particles and the $H_0$ hard particles have very different distributions while soft particles and $H_1$ hard particles have similar distributions. Bond-angle information therefore distinguishes between soft particles and $H_0$-type hard particles but not between soft particles and $H_1$-type hard particles. To fully distinguish between soft and hard particles, both radial and bond-angle information is needed.  Soft particles have environments that--at a minimum--have fewer particles in their nearest neighbor shell {\emph and} larger angles between adjacent particles. 

In summary, we have presented a novel ML method for identifying flow defects in disordered solids.  We note that we have focused on the short-time correlation of structure with particle rearrangements.   However, our method should shed light on the connection between local structural evolution and the correlation of rearrangements in time and space~\cite{keys11} over longer time scales in glassforming liquids.  We also note that we cannot predict the specific particles that will participate in rearrangements at a later time; rather, we identify a \emph{population} of particles that are likely to rearrange.  The latter quantity is more useful in thermal and/or sheared systems, since fluctuations lead to stochasticity in rearrangements.  

Our method relies on local structure alone, and can be applied directly to snapshots of experimental systems, in contrast to previous methods~\cite{manning11}.  Our approach also scales linearly with the number of particles, $N$, while vibrational mode approaches scale as $N^3$. The efficient identification of flow defects is critical to testing phenomenological approaches to plasticity based on flow defects~\cite{falk11,bocquet2009,mansard11,lin14}.   Previous applications of machine learning methods in physics have focused on approximation~\cite{GA_approx_hamil_zunger,NN_DFT_diamond,behlerPRL_orig} or on optimization and design tools~\cite{pred_crystal_struc_DM,pickard_Al,aspuru,highthroughput_highway}.   Our approach shows that such methods--designed for detecting subtle correlations--can also be used directly to gain conceptual understanding not achieved with conventional approaches.

\begin{acknowledgments}
E.D.C. and S.S.S. contributed equally to this work. We thank Amos Waterland, Carl Goodrich, Daniel Sussman and Franz Spaepen for helpful discussions. This work was supported by the UPENN MRSEC, NSF-DMR-1120901 (S.S.S., J.M.R.), NSF DMR-1305199 (D.J.D.), U.S. Department of Energy, Office of Basic Energy Sciences, Division of Materials Sciences and Engineering under Award DE-FG02-05ER46199 (A.J.L.), UBC Killam Faculty Research Fellowship (J.R.), and Harvard IACS Student Scholarship (E.D.C.). 
\end{acknowledgments}

\bibliography{bibliography.bib}

\end{document}